\documentclass[prl,twocolumn,superscriptaddress,showpacs,preprintnumbers,amsmath,amssymb,floats]{revtex4}
\usepackage{graphicx}
\begin{document}

\preprint{PRLv2s}
\title{Low-Energy Charge-Density Excitations in MgB$_{2}$: Striking
Interplay between Single-Particle and Collective Behavior for Large
Momenta}

\author{Y.~Q.~Cai}
\email{cai@nsrrc.org.tw}
\author{P.~C.~Chow}
\thanks{Present address: HP-CAT, Advanced Photon Source, Argonne
National Laboratory, Argonne, IL 60439, USA} \affiliation{National
Synchrotron Radiation Research Center, Hsinchu 30076, Taiwan}
\author{O.~D. Restrepo}
\affiliation{Department of Physics and Astronomy, The University of
Tennessee, Knoxville, Tennessee 37996-1200, USA}
\affiliation{Materials Science and Technology Division, Oak Ridge
National Laboratory, Oak Ridge, Tennessee 37831-6030, USA}
\author{Y.~Takano}
\author{K.~Togano}
\affiliation{National Institute for Materials Science, 1-2-1 Sengen,
Tsukuba 305-0047, Japan}
\author{H.~Kito}
\affiliation{National Institute of Advanced Industrial Science and
Technology, 1-1-4 Umezono, Tsukuba 305-8586, Japan}
\author{H.~Ishii}
\author{C.~C.~Chen}
\author{K.~S.~Liang}
\author{C.~T.~Chen}
\affiliation{National Synchrotron Radiation Research Center,
Hsinchu 30076, Taiwan}
\author{S.~Tsuda}
\affiliation{Institute for Solid State Physics, University of
Tokyo, Kashiwa, Chiba 277-8581, Japan}
\author{S.~Shin}
\affiliation{Institute for Solid State Physics, University of
Tokyo, Kashiwa, Chiba 277-8581, Japan} \affiliation{Institute of
Physical and Chemical Research (RIKEN), Sayo-gun, Hyogo 679-5198,
Japan}
\author{C.~C.~Kao}
\affiliation{National Synchrotron Light Source, Brookhaven National
Laboratory, Upton, NY 11973-5000, USA}
\author{W.~Ku}
\affiliation{Department of Physics, Brookhaven National Laboratory,
Upton, NY 11973-5000, USA}
\author{A.~G. Eguiluz}
\affiliation{Department of Physics and Astronomy, The University of
Tennessee, Knoxville, Tennessee 37996-1200, USA}
\affiliation{Materials Science and Technology Division, Oak Ridge
National Laboratory, Oak Ridge, Tennessee 37831-6030, USA}

\date{\today}

\begin{abstract}
A sharp feature in the charge-density excitation spectra of
single-crystal MgB$_{2}$, displaying a remarkable cosine-like,
periodic energy dispersion with momentum transfer ($q$) along the
$c^{*}$-axis, has been observed for the first time by
high-resolution non-resonant inelastic x-ray scattering (NIXS).
Time-dependent density-functional theory calculations show that the
physics underlying the NIXS data is strong coupling between
single-particle and collective degrees of freedom, mediated by large
crystal local-field effects. As a result, the small-$q$ collective
mode residing in the single-particle excitation gap of the B $\pi$
bands reappears periodically in higher Brillouin zones. The NIXS
data thus embody a novel signature of the layered electronic
structure of MgB$_{2}$.

\end{abstract}

%\keywords{}

\maketitle

The discovery of superconductivity in MgB$_{2}$ at a high critical
temperature $T_{c} \sim$ 39 K \cite{nagamatsu} has stimulated
numerous theoretical and experimental studies worldwide. Unlike
high-$T_{c}$ cuprates, the relatively simple crystal structure of
MgB$_{2}$ allows detailed first-principles calculations to be
performed and compared with experiments. As a result, MgB$_{2}$ is
now widely accepted as a phonon-mediated conventional superconductor
based on the anisotropic Eliashberg formalism with multiple gaps
facilitated by the different electron-phonon coupling strengths of
the boron $\sigma$ and $\pi$ bands
\cite{budko,an,kortus,liu,choi,tsuda}. The strong electron-phonon
coupling between the 2D $\sigma$ bands and the in-plane vibration of
the B layers (the $E_{2g}$ phonons) dominates the superconducting
properties and is largely responsible for the unusually high
$T_{c}$.

Within this phonon-mediated picture of superconductivity,
dynamically screened electron-electron and electron-ion interactions
play an important role \cite{mcmillan}. The study of charge-density
response function through inelastic x-ray \cite{schuelke} or
electron \cite{raether} scattering experiments and/or theoretical
calculations provides a powerful means for investigating the details
of the dielectric screening and the associated crystal potential,
local-field, and exchange-correlation effects. Recent
first-principles calculations \cite{moon,floris}, for example, have
shown the importance of the crystal local-field effects (CLFE) in
shaping the superconducting properties of MgB$_{2}$. Earlier
theoretical studies of the collective charge excitations in
MgB$_{2}$ based on first-principles calculations by Ku {\em et al.}
\cite{ku} and Zhukov {\em et al.} \cite{zhukov}, on the other hand,
predicted a sharp new collective mode between 2-5 eV for momentum
transfer $q {\parallel} c^{*}$-axis as a result of coherent charge
fluctuations between the Mg and B layers due to the unique
electronic structure of MgB$_{2}$. The feature as predicted
dispersed weakly with $q$ and decayed via Landau damping into the
single-particle continuum at $q \sim$ 6 nm$^{-1}$ in the first
Brillouin zone (BZ). The CLFE were reported to be negligible, for
the small $q$-range investigated.

In this Letter we demonstrate, using state-of-the-art
high-resolution non-resonant inelastic x-ray scattering (NIXS)
experiments and {\em ab initio} time-dependent density functional
theory (TDDFT) calculations, that the charge response of MgB$_{2}$
is truly remarkable. Indeed, the long-lived, low-energy collective
excitation in MgB$_{2}$ exists not only for small $q$ as previously
predicted \cite{ku,zhukov}, but it actually extends to higher BZ's
along the $c^{*}$-axis. The mechanism behind such physics is the
strong coupling between the single-particle and collective
excitation channels which is mediated by large crystal local-field
effects (CLFE) due to charge inhomogeneity normal to the Mg and B
layers. As a result, the conventional Landau-damping mechanism does
not restrict the collective mode to a small fraction of the momentum
space: the mode actually reappears {\em periodically} as $q$ steps
through successive BZ's. The enhanced scattering cross section at
high $q$ and the negligible multiple-scattering effects of the x-ray
measurements play a decisive role for the clean observation of this
mode over a large momentum range. Our work represents also the first
observation of collective charge excitations of this kind in
condensed matter physics \cite{note0}.

The MgB$_{2}$ single crystals were prepared by high-pressure
sintering of MgB$_{2}$ powder \cite{takano}. Superconducting
transition with the onset temperature of 39 K was confirmed by both
magnetic and resistive measurements. The crystals were plate-like,
golden in color, and measured $\sim500\times300\times20$
${\mu}$m$^{3}$ in size. The crystal quality and orientation were
characterized by Laue and indexing rotation photos, in which
well-defined principal Bragg diffraction spots were identified and
used to align the $c^{*}$- and $a^{*}$-axis to the scattering plane.
Crystals chosen for the measurement showed rocking curve widths of
$\sim 0.05^{\circ}$ for low order reflections. Beam damages to the
crystals were not visible upon visual inspection using a microscope
before and after the measurements.

NIXS spectra were collected at room temperature on the Taiwan
inelastic x-ray scattering beamline BL12XU at SPring-8 \cite{cai}.
The energy transfer was varied by scanning the incident energy
relative to the near backscattering energy (9.886 keV) of a 2-m
radius Si(555) spherical analyzer \cite{wang}. The total energy
resolution was 65 and 250 meV respectively using two configurations
of the beamline and the spectrometer \cite{cai}. The momentum
resolution was 0.6 nm$^{-1}$ in the horizontal scattering plane, and
2.3 nm$^{-1}$ in the vertical plane.

\begin{figure}[b]
\scalebox{1.2}{\includegraphics*[0.55cm,0.55cm][8cm,5.5cm]{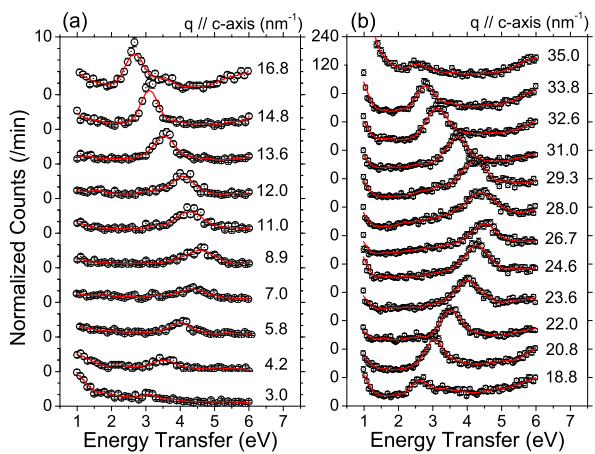}}
\caption{\label{fig1} NIXS spectra at various momentum transfers $q
\parallel c^{*}$-axis showing the low-energy collective mode, where
$q = 8.9$ nm$^{-1}$ corresponds to the first boundary of the
extended Brillouin zones (BZ). The total energy resolution was 65
meV for (a), and 250 meV for (b).}
\end{figure}

\begin{figure}[t]
\scalebox{0.7}{\includegraphics*[0.5cm,0.5cm][12.5cm,8.5cm]{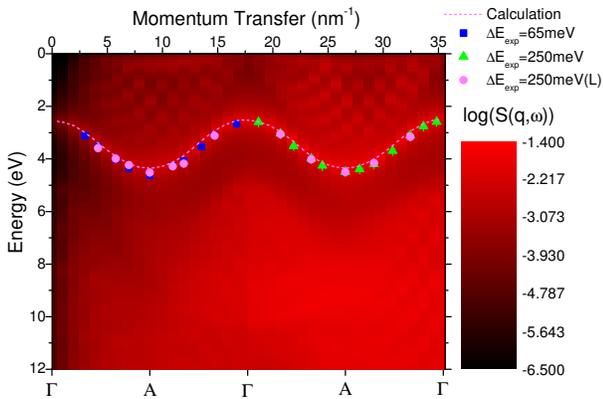}}
\caption{\label{fig2} (Color) Theoretical $S(\mathbf{q},\omega)$
calculated in the present work in false color log scale as a
function of energy and momentum transfer showing the cosine energy
dispersion of the low-energy collective mode. Filled squares and
triangles mark the energy positions obtained from the NIXS spectra
shown in Fig.~\protect\ref{fig1}, whereas filled circles are data
from another set of spectra taken with a total energy resolution of
250 meV. }
\end{figure}

Figure~\ref{fig1} shows a selection of the NIXS spectra taken over
the energy region of the low-energy collective mode with momentum
transfer $q$ covering four entire BZ's along the $c^{*}$-axis. All
spectra are raw data normalized to the incident beam intensity in
unit of counts per minute. The periodic nature of this mode in
energy is apparent: The feature starts at $\sim 3$ eV at $q = 3.0$
nm$^{-1}$ [Fig.~\ref{fig1}(a)], disperses down to $\sim 4.5$ eV at
the first zone boundary at 8.9 nm$^{-1}$ where the dispersion turns
negative until it reaches near the next zone center at 17.8
nm$^{-1}$, and then repeats itself in the third and the fourth BZ
[Fig.~\ref{fig1}(b)]. The width of the loss feature varies also
periodically, with about 0.5-eV (1-eV) width at half maximum near
the zone centers (boundaries). Experimentally, the energy dispersion
of this mode with momentum transfer can be described entirely by a
simple cosine function (see Fig.~\ref{fig2}): $\omega = {\omega}_{0}
- 2{\gamma}{\rm cos}(qc)$, with ${\omega}_{0}=3.55$ eV,
${\gamma}=0.49$ eV, and $c=0.352$ nm the lattice constant of
MgB$_{2}$ along the $c$-axis. An extrapolated value of 2.57 eV at $q
= 0$ can thus be obtained, which is in excellent agreement with
recent optical studies \cite{guritanu} and provides additional
support for the experimental confirmation of this collective mode at
low $q$.

\begin{figure}[]
\scalebox{1.3}{\includegraphics*[0.2cm,0.2cm][8cm,6.5cm]{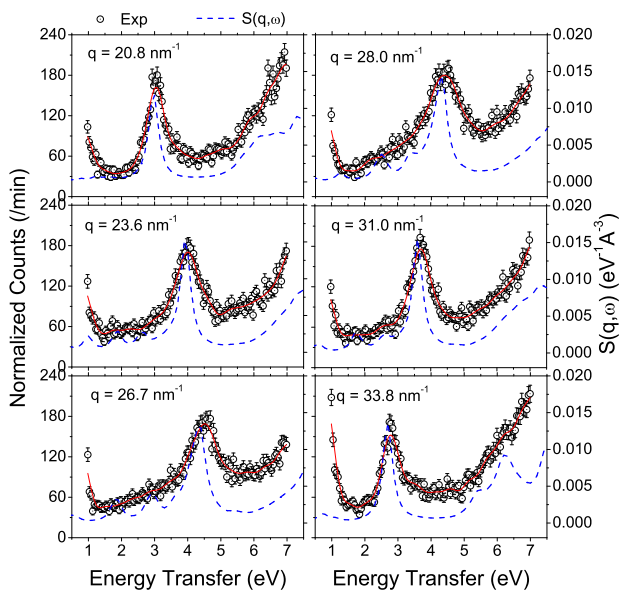}}
\caption{\label{fig3} Qualitative comparison between the calculated
$S(\mathbf{q},\omega)$ and the NIXS spectra at selected $q$ values
of the second period.}
\end{figure}

In order to understand the physics behind the data presented in
Fig.~\ref{fig1}, we have calculated the dynamical structure factor
$S(\mathbf{q},\omega)$ based on time-dependent density functional
theory (TDDFT) \cite{petersilka}, including fully the CLFE. Being a
quantity directly proportional to the scattering cross section of
NIXS \cite{schuelke}, the dynamical structure factor
$S(\mathbf{q},\omega)$ is related to the density-response function
$\chi$ by the fluctuation-dissipation theorem,
\begin{equation}\label{eqn1}
    S(\mathbf{q},\omega) = -2{\hbar}V{\rm
Im}{\chi}_{\mathbf{G_{q}},\mathbf{G_{q}}}(\mathbf{q}-\mathbf{G_{q}},\omega),
\end{equation}
where $\mathbf{G_{q}}$ is the unique vector of the reciprocal
lattice which brings $\mathbf{q}$ into the first BZ. In TDDFT,
$\chi$ obeys the formally exact integral equation $\chi = {\chi}_{S}
+ {\chi}_{S} (\nu + f_{XC}){\chi}$, where ${\chi}_{S}$ is the
response function for Kohn-Sham electron-hole pairs, $\nu$ the
Coulomb interaction, and $f_{XC}$ the many-body kernel
\cite{petersilka}. ${\chi}_{S}$ is evaluated for a ground state
obtained in the local-density approximation (LDA) using the LAPW
method \cite{blaha}. The many-body kernel $f_{XC}$ is evaluated for
the adiabatic extension of the LDA \cite{gross}, defining the TDLDA
response, which in view of our NIXS data, is an excellent
approximation \cite{note2}. The integral equation is then solved for
the response matrix
${\chi}_{\mathbf{G},\mathbf{G}^{\prime}}(\mathbf{q}-\mathbf{G_{q}},\omega)$
 \cite{gurtubay}. The calculated
$S(\mathbf{q},\omega)$, obtained using a damping parameter of 0.17
eV in the evaluation of ${\chi}_{S}$, is compared with experiment in
Figs.~\ref{fig2} and \ref{fig3}.

The calculated periodic energy dispersion of the collective mode
represented by the dashed line in Fig.~\ref{fig2} can be seen to
agree almost perfectly with experiment throughout the four BZ's
investigated. In Fig.~\ref{fig3}, the calculated
$S(\mathbf{q},\omega)$ is compared in detail with the NIXS spectra
at a few selected $q$ values of the second period. Here no attempt
has been made to convert the experimental raw data onto absolute
units due to insufficient data range to apply the f-sum rule
\cite{tischler}. The comparison is made therefore qualitatively with
a constant scale factor between theory and experiment for all the
$q$ values shown. Nevertheless, the spectral behavior of the
observed collective mode is reproduced very well by the calculation
despite an overall underestimate of the spectral weight at the
higher energy side.

Now, the density-response matrix can be expressed in terms of the
inverse of the dielectric matrix as
${\chi}_{\mathbf{G},\mathbf{G}^{\prime}}(\mathbf{q}-\mathbf{G_{q}},\omega)
= {\nu}^{-1}(\mathbf{q} - \mathbf{G_{q}} + \mathbf{G})
\left\{\left[{\epsilon}(\mathbf{q} -
\mathbf{G_{q}},\omega)\right]_{\mathbf{G},\mathbf{G}^{\prime}}^{-1}
- {\delta}_{\mathbf{G},\mathbf{G}^{\prime}}\right\}$. For the
latter, we have derived the following exact result \cite{restrepo},
\begin{eqnarray}\label{eqn2}
\left[{\epsilon}(\mathbf{q} -
\mathbf{G_{q}},\omega)\right]_{\mathbf{G_{q}},\mathbf{G_{q}}}^{-1} =
\frac{1}{{\epsilon}_{\mathbf{G_{q}},\mathbf{G_{q}}}(\mathbf{q}-\mathbf{G_{q}},\omega)}
\nonumber \\
+ F(\mathbf{q},\mathbf{q} -
\mathbf{G_{q}};\omega)\left[{\epsilon}(\mathbf{q} -
\mathbf{G_{q}},\omega)\right]_{\mathbf{0},\mathbf{0}}^{-1}
\end{eqnarray}
for $\mathbf{q}$'s in higher BZ's \cite{note1}. Here
$F(\mathbf{q},\mathbf{q}-\mathbf{G_{q}};\omega) = -
W(\mathbf{q},\mathbf{q}-\mathbf{G_{q}};\omega)/{\epsilon}_{\mathbf{G_{q}},
\mathbf{G_{q}}}(\mathbf{q} - \mathbf{G_{q}},\omega)$, where
$W(\mathbf{q},\mathbf{q} - \mathbf{G_{q}};\omega) =
\left(Det\left[\epsilon\right] -
\epsilon_{\mathbf{G_{q}},\mathbf{G_{q}}}M_{\mathbf{G_{q}},
\mathbf{G_{q}}}\right)/M_{\mathbf{0},\mathbf{0}}$,
$M_{\mathbf{0},\mathbf{0}}$ and $M_{\mathbf{G_{q}},\mathbf{G_{q}}}$
being the minor of the
($\mathbf{G}=\mathbf{0},\mathbf{G^{\prime}}=\mathbf{0}$) and
($\mathbf{G}=\mathbf{G_{q}},\mathbf{G^{\prime}}=\mathbf{G_{q}}$)
elements of $\epsilon$, respectively. Here, the arguments of all
quantities shown symbolically are
($\mathbf{q}-\mathbf{G_{q}},\omega$).

Equation (\ref{eqn2}) brings out the physics of the charge response
for $\mathbf{q}$'s in higher BZ's quite vividly. The first term
corresponds to the response appropriate for homogeneous media; it
gives rise to charge fluctuations involving incoherent electron-hole
pairs. The second term introduces all the CLFE. For $\mathbf{q} -
\mathbf{G_{q}}$ near the zone center, collective charge fluctuations
(plasmons) are built into the response function
$\left[{\epsilon}(\mathbf{q} -
\mathbf{G_{q}},{\omega})\right]_{\mathbf{0},\mathbf{0}}^{-1}$;
whether such excitation is actually realized in
$S(\mathbf{q},\omega)$ (via Eq. (\ref{eqn1})) depends on the impact
of the {\em material-dependent} coupling function $F(\mathbf{q},
\mathbf{q} - \mathbf{G_{q}},\omega)$.

\begin{figure}[b]
\scalebox{0.9}{\includegraphics*[0cm,1.9cm][7cm,7.4cm]{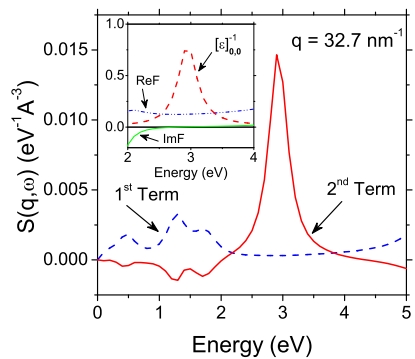}}
\caption{\label{fig4} Contributions calculated from the first and
second terms in Eq.~(\ref{eqn2}) to the $S(\mathbf{q},\omega)$ of
MgB$_{2}$ at $q=32.7$ nm$^{-1}$ along the $c^{*}$-axis. The insert
shows the real and imaginary parts of the $F$ factor and the
$\left[{\epsilon}\right]^{-1}_{\mathbf{0},\mathbf{0}}$ for the
corresponding $\left|\mathbf{q}-\mathbf{G_{q}}\right| = 2.9$
nm$^{-1}$, where $\mathbf{G_{q}}=(2\pi/c)(0,0,2)$.}
\end{figure}

In Fig.~\ref{fig4}, the contributions to the $S(\mathbf{q},\omega)$
of MgB$_{2}$ from the first and second terms of Eq.~(\ref{eqn2}) are
shown for $q=32.7$ nm$^{-1}$ in the fourth BZ along the
$c^{*}$-axis. Clearly, in the energy region relevant for the
collective mode associated with the single-particle excitation gap
of the B $\pi$ bands \cite{ku,zhukov}, the first term amounts to a
weak background, while the second term features a sharp peak, which
is typical of $\mathbf{q} - \mathbf{G_{q}}$ near the zone center
($\left|\mathbf{q}-\mathbf{G_{q}}\right| = 2.9$ nm$^{-1}$ in
Fig.~\ref{fig4}). Now as shown in the insert of Fig.~\ref{fig4},
${\rm Re}F$ is {\em positive and nearly constant} for energies
around the collective mode, while ${\rm Im}F$ goes through a local
zero. Such an $F$ ensures that the second term of Eq.~(\ref{eqn2})
is controlled by ${\rm Im}\left[{\epsilon}(\mathbf{q} -
\mathbf{G_{q}},{\omega})\right]_{\mathbf{0},\mathbf{0}}^{-1}$. In
other words, in MgB$_{2}$ $F$ feeds the small-$(\mathbf{q} -
\mathbf{G_{q}})$ physics into the charge fluctuations for large
$\mathbf{q}$'s. The plasmon can be viewed as acting - via $F$ - as a
source (additional to the external field of the x-ray photon)
driving a collective charge fluctuation generated in a NIXS event
for large $\mathbf{q}$'s. It is important to notice that this
process is intrinsically periodic. Indeed, the fact that for higher
BZ's the loss feature retraces its energy dispersion in the first BZ
reflects the impact on $S(\mathbf{q},\omega)$, via the coupling
function $F$ in Eq. (\ref{eqn2}), of the ``periodic'' condition
${\epsilon}_{\mathbf{G},\mathbf{G}^{\prime}}(\mathbf{q} -
\mathbf{G_{q}},\omega) =
{\epsilon}_{\mathbf{G},\mathbf{G}^{\prime}}(\mathbf{k},\omega)$,
where $\mathbf{k}$ is in the first BZ.

It is also interesting to note that there is a non-trivial
difference between the response in the higher BZ's and that in the
first one. In higher BZ's the collective mode arises from the
``propagation'' of the excitation gap of the B $\pi$ bands built
into ${\epsilon}_{\mathbf{0},\mathbf{0}}$ via the inverse dielectric
matrix of the second term in Eq. (\ref{eqn2}). {\em This is the
essence of the CLFE, in this context}. By contrast, in the first BZ
the excitation gap is directly built into the scalar response
$1/{\epsilon}_{\mathbf{0},\mathbf{0}}(\mathbf{q},\omega)$
\cite{note1}.

Additional insight into the charge response of MgB$_{2}$ is achieved
in light of a simplified version of Eq.~(\ref{eqn2}), corresponding
to the $2{\times}2$ dielectric-matrix model invoked by Sturm,
Sch\"{u}lke, and Schmitz \cite{sturm} in their study of the
$S(\mathbf{q},\omega)$ of Si. For Si, $F$ is complex with a {\em
negative} real part; thus the CLFE yield only a {\em weak Fano-type}
resonance, which highlights the novelty of the charge response of
MgB$_{2}$. As we show elsewhere \cite{restrepo}, a $2{\times}2$
dielectric-matrix model represents a fair approximation to the full
calculations we have performed for MgB$_{2}$ using dielectric
matrices of rank up to 23 and an $18{\times}18{\times18}$ $k$-mesh.
(For $\mathbf{q} - \mathbf{G_{q}}$ near each higher BZ boundary a
$3{\times}3$ model becomes necessary \cite{restrepo}.) In such a
$2{\times}2$ model, the response function from the second term of
Eq.~(\ref{eqn2}) becomes simply
$1/{\epsilon}_{\mathbf{0},\mathbf{0}}(\mathbf{q}-\mathbf{G_{q}},\omega)$,
which makes obvious the identification of the plasmon discussed
above with the collective mode discussed in Refs. \cite{ku} and
\cite{zhukov} for $\mathbf{q}$'s near the first BZ $\Gamma$ point.
Most importantly, the fact that a dielectric matrix of rank two (or
three) describes the response of MgB$_{2}$ rather accurately is
traced to the layered structure of MgB$_{2}$, which yields a first
shell of non-zero $\mathbf{G}$'s containing only two vectors, and to
the fact that the length scale of the polarization in real space is
determined by orbitals which are rather extended, leading to a
sufficiently fast decay of
${\epsilon}_{\mathbf{G},\mathbf{G}^{\prime}}$; therefore, increasing
the rank of the $\epsilon$-matrix beyond two or three does not
affect the physics, in contrast to the case of, e.g.,
transition-metal oxides \cite{restrepo}.

In conclusion, the electron-hole degrees of freedom in MgB$_{2}$
lead to a novel charge response: long-lived {\em collective}
charge-density fluctuations can be excited in a NIXS event involving
large momenta - corresponding to length scales which are comparable
with the size of the B p orbitals. This is qualitatively different
from the case of simple metals and covalent semiconductors discussed
in Ref.~\cite{sturm}, the difference stemming from the fact that the
coupling function $F$ is $\sim 60$ times larger in MgB$_{2}$. The
CLFE were shown to dominate the physics of the charge excitations
for large momenta, leading to the striking periodicity exhibited by
the observed low-energy excitation as the momentum transfer steps
through successive BZ's. Overall, the impact of the CLFE on the
charge response of MgB$_{2}$ rivals their importance in materials
involving confined geometries such as carbon nanotubes
\cite{marinopoulos} and superlattices \cite{botti}. Ultimately, the
nature of the measured charge excitations stems from the layered
electronic structure of MgB$_{2}$ and the delocalized nature of the
orbitals involved in the screening. Analogous physics should be at
play in other layered compounds of current interest.

\begin{acknowledgments}

Stimulating discussions with B. Friedman, B.~C. Larson, J.~Z.
Tischler, A. Shukla, M. Calandra, and K.-D. Tsuei are acknowledged.
A.G.E. acknowledges support from NSF ITR DMR-0219332. Research at
ORNL is sponsored by the DOE, Office of Science, DMS under contract
with UT-Battelle, LLC. W.K. and A.G.E. acknowledge collaboration via
the DOE/CMSN. This work was carried out at SPring-8 under
experiments No. C03A12XU-1500N and C03B12XU-1503N, and was partly
supported by the National Science Council of Taiwan.

\end{acknowledgments}


\begin{references}
\bibitem{note2}
As it turns out, in MgB$_2$ the many-body kernel plays a small role;
in fact, even ignoring it altogether yields spectra which are in
good agreement with our IXS data.
\end{references}
\end{document}